\documentclass{svproc}
\usepackage{url}
\usepackage[pdftex]{graphicx}
\usepackage{amsmath}

\begin{document}
\mainmatter              
\title{Simulation method for evaporative cooling of trapped Bose gases at finite temperatures}
%
%
\author{Emiko Arahata\inst{1} \and Tetsuro Nikuni\inst{2}}
%
%
%
\institute{Tokyo Metropolitan University, 1-1 Minamiosawa, Hachioji-shi, Tokyo, Japan,\\
\email{arahata@tmu.ac.jp}
\and
Tokyo University of Science, 1-3 Kagurazaka, Shinjuku, Tokyo, Japan}

\maketitle              

\begin{abstract}
We develop a simulation method for evaporative cooling of trapped Bose-Einstein condensate at finite temperatures using Zaremba-Nikuni-Griffin (ZNG) formalism. ZNG formalism includes the generalized GP equation and a semiclassical kinetic equation for the thermal cloud, which treats the excitations semiclassically within the Hartree Fock approximation. The generalized GP equation includes the mean field due to the thermal cloud and the source term associated with collisions between the condensate and the thermal cloud. 
Our method is based on the numerical approach developed by Jackson and Zaremba, which simulates the kinetic equation using test particles. A key point of our method is to mimic the evaporative cooling process by eliminating the test particles with high energy. We show that our method successfully describes condensate growth during evaporative cooling. We also numerically simulate vortex lattice formation during evaporative cooling in the presence of the rotating thermal cloud. 
\end{abstract}
\section{Introduction}\label{sec1}
Experimentally, atomic Bose-Einstein condensates (BEC) are created by
evaporatively cooling trapped Bose atoms.
The evaporative cooling technique consists of the selective removal of high-energy atoms and collisional relaxation of the remaining gases.
This is the finite-temperature dynamics in which the condensation and the thermal cloud coexist.
The experiments at finite temperatures have observed many phenomena that can be attributed to the existence of the thermal cloud, such as vortex-lattice creation and Landau damping.
In particular, an interesting experiment of vortex-lattice creation was reported in 2001 by the JILA group \cite{PhysRevLett.87.210403}.
This experiment created vortex lattices by first spinning a normal Bose gas and then cooling the rotating gas through the BEC transition. 
This experiment indicated that vortex nucleation can arise from angular momentum transfer between the thermal cloud and the condensate.
Ref.~ \cite{PhysRevLett.87.210403} also showed that 
the above procedure can create vortex lattices 
more efficiently than creating BEC and then 
rotating it.

Motivated by the experiment of Ref.~\cite{PhysRevLett.87.210403}, in this paper, we study the dynamics of evaporative cooling of rotating Bose gases at finite temperatures.
For this purpose, we first develop a simulation 
method for evaporative cooling. 
The dynamics of evaporative cooling has been
theoretically studied by using the quantum kinetic theory within the so-called  
``ergodic approximation''
\cite{C_W_Gardiner_PRL82,M_J_Bijlsma_PRA62}.
However, in order to clarify the dynamical effect of the thermal cloud, it is important to simulate the full dynamics of the thermal cloud. 
In this paper, we use the formalism of Zaremba, Nikuni, and Griffin (ZNG) to include the generalized GP equation and a semiclassical Boltzmann equation for the thermal cloud, which treats the excitations semiclassically within the Hartree Fock (HF) approximation.
The generalized GP equation includes the mean
field due to thermal could and source term associated with collisions between the condensate and the thermal cloud.
We note that in Ref\cite{E_arahata_JLTP2016}, we
simulated the ZNG equations in the presence of a rotating trap potential, and showed that 
vortex creation occurs due to the transfer of angular momentum between the condensate and the thermal cloud. 
However, vortex creation by evaporative
cooling of rotating gas has not been 
studied theoretically.

The paper is organized as follows. In Sec. 2, we summarize ZNG equations.  In Sec. 3, we explain our approach to evaporative cooling. In Sec. 4, we discuss the evaporative cooling of a rotating Bose gas.
Finally, we conclude in Sec. 5.

\section{ZNG equations for a trapped Bose-condensed gas}\label{sec2}

The ZNG formalism consists of a generalized GP equation for the condensate order parameter $\Phi({\bf r},t)$ and  
the semiclassical Boltzmann kinetic equation for the
thermal cloud distribution function $f({\bf p},{\bf r},t)$. 
The derivation of the ZNG formalism can be found in Ref.~\cite{T_ZNG_JLTP}
 The coupled ZNG equations are given by
  \begin{eqnarray}
&&i\hbar \frac{\partial \Phi({\bf r},t)}{\partial t}
=\biggl[-\frac{\hbar^2 \nabla^2}{2m}+V_{\rm ext}({\bf r},t)+gn_{\rm c}({\bf r},t)+2g\tilde n({\bf r},t)
\nonumber\\
&& \hspace{2.5cm}
-iR({\bf r},t) \biggr]\Phi({\bf r},t),
\label{eq_6}\\
&&\frac{\partial f ({\bf p},{\bf r},t)}{\partial t }+\frac{{\bf p}}{m}\cdot \nabla f ({\bf p},{\bf r},t) - \nabla U \cdot \nabla_{\bf p} f({\bf p},{\bf r},t)
    =C_{12}[f]+C_{22}[f].\label{eq_7}
      \end{eqnarray}
where $n_{\rm c}=| \Phi |^2$ is the condensate density, $\tilde n=\int \frac{d{\bf p}}{(2\pi\hbar)^3}f({\bf p,r},t)$ is the noncondensate density, and $V_{\rm ext}({\bf r},t)$ is a time-dependent trap potential.
The $C_{22}$ term 
describes two-body collisions between noncondensate atoms,
and the  $C_{12}$ term describes collisions
between condensate and noncondensate atoms.
The detailed expressions for the $C_{22}$ and $C_{12}$ terms are given in Ref.~\cite
{T_ZNG_JLTP}.
In Eq.~(\ref{eq_7}), $U=V_{\rm ext}+2gn_{\rm c}+2g\tilde n$ is the effective potential for the noncondensate, which includes the self-consistent Hartree--Fock mean field. 
In the ZNG model, noncondensate atoms have the 
the local energy
 $
        \tilde \epsilon_p({\bf r},t)=p^2/(2m) +U({\bf r},t)    \label{eq_le}
$,
and the condensate has the local energy
$\epsilon_c=\mu_c+\frac{1}{2}mv_c^2$,
where 
${\bf v}_c\equiv -\hbar (\Phi^*\nabla\Phi-\Phi\nabla\Phi^*)/(2imn_c)$ and
$\mu_c\equiv \hbar^2 \nabla ^2 \sqrt{n_c}/(2m \sqrt{n_c})
+V_{\rm ext} 
+gn_c+2g\tilde n$
is the local condensate chemical potential.
    The source term $R$ appearing in Eq.~(\ref{eq_6}) is directly related to the $C_{12}$ collision term 
 through
%
$R =(\hbar/2n_c)\int d{\bf p}/(2\pi \hbar)^3C_{12}$.
   
     The numerical procedure for solving ZNG equations was developed in Ref. \cite{PhysRevA.66.033606}. 
     In this approach, the dynamics of the thermal cloud is calculated by using $N$-body simulations. The dynamics of the condensate is determined by numerically propagating the GP equation using a split-operator fast Fourier transform method.
     We use this approach to simulate evaporative cooling.

\section{Simulation method for evaporative cooling}\label{sec3}

Before explaining our approach to evaporative cooling, we briefly mention the earlier studies in 
Refs.\cite{C_W_Gardiner_PRL82,M_J_Bijlsma_PRA62}.
Theoretical descriptions of the evaporative cooling 
in Refs.\cite{C_W_Gardiner_PRL82,M_J_Bijlsma_PRA62} are based on the ergodic approximation, in which the noncondensate distribution function is given as a function of the single-particle energy.
This approximation allows one to derive a simplified kinetic equation for the distribution function.
The evaporative cooling process is modeled by using a Bose distribution function truncated at a cutoff 
energy as an initial state.
Albeit approximate, this approach provides a physically reasonable description of evaporative cooling.

In the present study, we propose a simulation method for evaporative cooling based on the 
Jackson-Zaremba approach \cite{PhysRevA.66.033606}.
The main difficulty of applying this approach to evaporative cooling is that one does not directly 
work with the distribution function.
Instead, one solves the classical $N$-body problem for test particles and constructs the phase-space distribution function.
In this case, it is not obvious how to incorporate the evaporation process into the simulation.
In the present study, we model evaporative cooling by
eliminating the test particles with energy above a cutoff energy.
One can then effectively prepare a truncated distribution.

More explicitly,
we simulate the time evolution of test particles following the procedure described
in Ref.~\cite{PhysRevA.66.033606}.
After updating the phase-space coordinates $({\bf r}_i,{\bf p}_i)$ of
test particles at each timestep 
  of calculating the collisional term,
we eliminate the test particle that has the single-particle energy greater than the cutoff energy:
 $\tilde \epsilon_{p_i}({\bf r}_i,t)>\epsilon^{{\rm max}}_{\rm cut}(
  t)$.
   In order to describe the evaporative cooling, cutoff energy must be lowered in time.
  We adopt the following time dependence of the cutoff energy:
$
  \epsilon_{\rm cut}(t)=\epsilon^{0}_{\rm cut}\exp(-\alpha t^2) .
$
  In order to determine the initial cutoff energy 
$\epsilon^{{\rm max}}_{0}$,
we first find the maximum single-particle
energy $\tilde \epsilon^{\rm {max}_0}_{p}({\bf r},0)=\max \tilde \epsilon_{p_i}({\bf r}_i,t=0)$
in the initial state $t=0$.
We then set the initial cutoff energy as
$\epsilon_{\rm cut}^0=\gamma\tilde \epsilon^{{\rm max}_0}_{p}$.
The optimal values for 
$\alpha$ and $\gamma$ have been chosen
such that evaporative cooling works the most efficiently.
We varied the value of $\alpha$ in increments of 0.01 and the value of $\gamma$ from $0.5$ to $0.9$ in increments of 0.05 and
observed the cooling behavior. We found that the values $\alpha=0.01$ and $\gamma=0.85$ resulted in the fastest attainment of a temperature of approximately 0.5 times the BEC transition temperature ($T\sim 0.5T_c$).
That is, if the maximum cutoff energy is too large the evaporation does not occur.
On the other hand, if the maximum cutoff energy is too small, the system quickly loses the particles and one cannot retain the phase-space density.
 As for the time dependence of $\epsilon_{\rm cut}(t)$, 
we explored the other monotonically decreasing functions, such as 
linear, quadratic, and logarithmic functions.
However, we found that the exponential function allows for the fastest attainment of $T\sim 0.5T_c$ without any numerical uncertainty.

\begin{figure}[htbp]
  \includegraphics[height=1.3in]{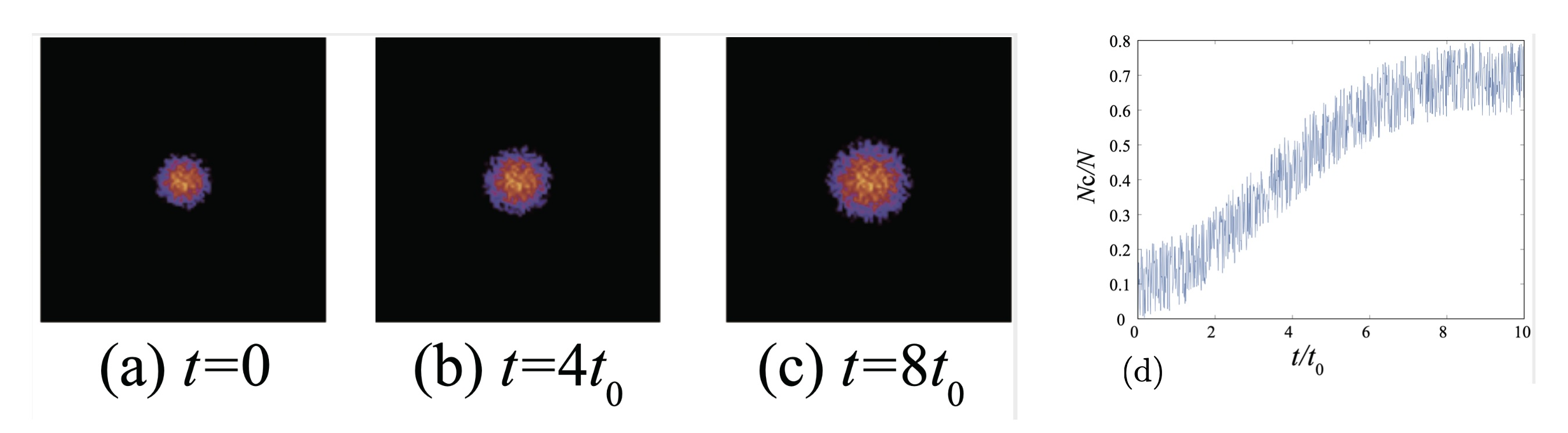}
 \caption{Time evolution of the condensate. The
 two-dimensional condensate density sliced at $x=0$ at the time (a)$t=0$ (b) $t=4t_0$ (c) $t=8t_0$.
 Panel (d) shows the time evolution of the condensate fraction $N_{\rm c}/N$.} 
  \label{Fig:evt08}
\end{figure}

In the numerical simulation shown below, we set the parameters of the system so as to make the ZNG equations in a rotating trap potential remarkably stable: a total of $N_0=6\times 10^6$ $^{87}$Rb atoms,
a radial trap frequency $\omega_{\rho}/2\pi=6.8$Hz 
and the axial tap frequency $\omega_{z}=13.6$.
The simulation range is $10R_{\rm TF}$, where
$R_{\rm TF}= a_{\rm ho}\left(8\pi Na/a_{\rm ho}\right)^{1/5}$with $a_{\rm ho}=\sqrt{\hbar/m\omega_\rho}$ is the 
Thomas-Fermi radius,
and the spatial computational mesh for the condensate is $\sim0.1\xi$ (where $\xi=\hbar/\sqrt{2ngn_0}$ is the healing length).
Finally, we choose the total number of test particles as $N_{\rm test}=1.0 \times 10^8$.
We prepare an initial equilibrium state at $T\sim 0.95 T_{\rm c}$ \cite{E_arahata_JLTP2016}. 
Figure \ref{Fig:evt08} shows the simulation results for the time evolution of the condensate
during the evaporative cooling. 
Figures \ref{Fig:evt08}(a)-(c) display the two-dimensional condensate densities sliced at $x=0$ from the complete three-dimensional data.
The initial equilibrium density distribution at $t=0$ is shown in Fig. \ref{Fig:evt08}(a). The evaporate cooling is started at $t=0$.
Fig.~\ref{Fig:evt08}(b) and Fig.~\ref{Fig:evt08}(c) show the growth of the condensate during evaporative cooling. 
In Fig. \ref{Fig:evt08}(d), we plot the time evolution of the condensate fraction $N_{\rm c}/N$. We can see that $N_{\rm c}/N$ increases as time evolves. 
From the condensate fraction, the final temperature
at $t=t_0$ is estimated as $T\simeq 0.56T_c$

The pronounced oscillations in the behavior of the condensate fraction can be attributed to the fact that the ZNG simulation takes into account collisions between particles in the condensate and the noncondensed component. These oscillations arise from processes where the condensate transitions into the noncondensed component and vice versa.
We confirmed that our procedure successfully 
simulates the growth of the condensate by evaporative cooling.

\section{Evaporative cooling of a rotating Bose gas}
In this section, we discuss the evaporative cooling of a rotating Bose gas.
In the JILA experiment,
vortex lattices were created by first spinning a normal Bose gas and then
cooling the rotating gas through the BEC transition. 
However, the ZNG formalism does not allow one to go through the BEC transition since this formalism is based on the mean-field description.
Therefore, we start with the temperature below, but very close to the BEC transition temperature where the condensate fraction is very small.

The system parameters are the same as in the previous section, except that we now 
consider a weakly anisotropic potential rotating with time-dependent rotation frequency $\Omega(t)$:
\begin{eqnarray}
V_{\rm ext}({\bf r},t)&=&\frac{m}{2}\omega_{\rm rad}^2 \left[(1+\epsilon)x^{'2}+
 (1-\epsilon)y^{'2} \right]
 +\frac{m}{2}\omega_{z}^2 z^2,
 \label{Eq:trap}
\end{eqnarray}   
where 
$
x'=x\cos[\Omega(t) t]+y\sin[\Omega(t) t], \
y'=-x\sin[\Omega(t) t]+y\cos[\Omega(t) t].$
We first prepare the
initial equilibrium state at $T=0.9T_{\rm c}$
in the static potential $V_{\rm ext}({\bf r},0)$.
We then start rotating the anisotropic trap potential
at $t=0$, gradually increasing the rotation frequency $\Omega(t)$.
More explicitly, we change the rotation frequency according to 
\begin{align}
    \Omega(t)= \left\{\begin{array}{ll}\Omega_{\rm max} \exp[-a(t-t_{\rm in})^2] ~~~~~~~~~~& (t<t_{\rm in}) \\
    \Omega_{\rm max} & (t_{\rm in}<t<t_{\rm out})  \end{array}\right.
    \label{eq:4}
	\end{align}
 Throughout this paper, we set $\Omega_{\rm max}=0.8 \omega_{\rm rad}$, $\epsilon=0.001$, $t_{\rm in}=t_0$, $t_{\rm out}=3t_0$, and $a=10/t_0^2$ with $t_0=10/\omega_{\rm rad}$. 
 With the present choice of parameters, we effectively start with $\Omega(0)\approx 0$.
We stop increasing the rotation frequency at $t=2t_0$.
At $t=4t_0$, where the system reaches a stationary state, we stop rotating the trap potential and start evaporative cooling. 
Finally, we stop evaporative cooling at $t=8t_0$.

\begin{figure}[htbp]
  \includegraphics[height=2.7in]{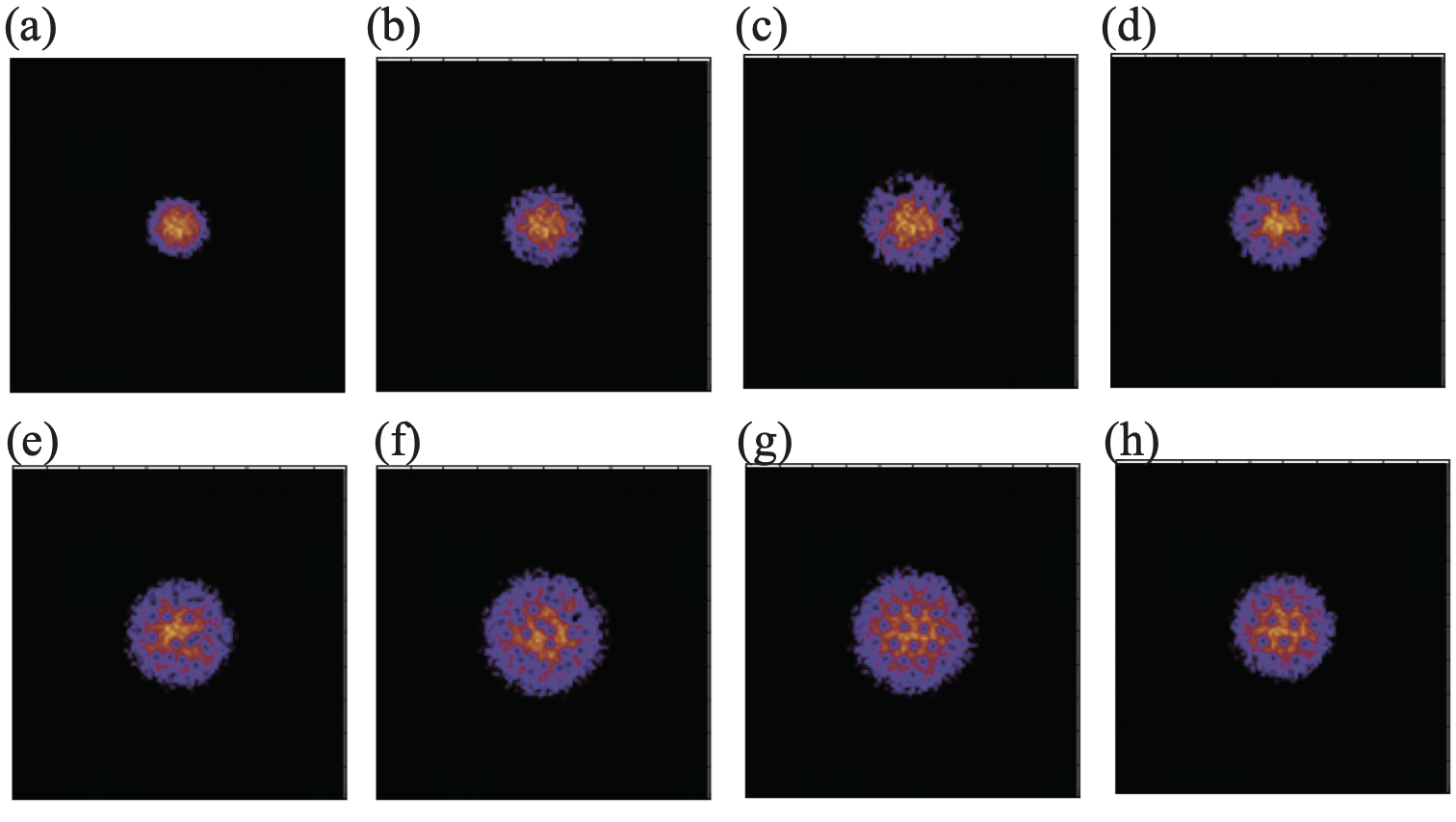}
 \caption{Time evolution of the condensate density profile in the rotating frame. The times are (a)$t=0$ (b)$t=4t_0$(c)$t=5t_0$(d) $t=6t_0$(e) $t=7t_0$ (f) $t=8t_0$ (g)$t=10t_0$ (h)$t=15t_0$.} 
  \label{Fig:t0-5}
\end{figure}

In Fig.\ref{Fig:t0-5}, we show the evolution of the condensate density during the evaporative cooling.
 The initial equilibrium density distribution at $t=0$ is shown in Fig. \ref{Fig:t0-5}(a). Figure \ref{Fig:t0-5}(b) shows the equilibrium density distribution in the presence of a rotating thermal cloud. 
We start the evaporate cooling at $t=4t_0$ and the time evolution of the condensate density profile during evaporate cooling is shown in Figs \ref{Fig:t0-5}(c)-(f). We can see the growth of condensate as well as the excitation of surface mode. The vortices then begin to penetrate into the condensate. Figure \ref{Fig:t0-5}(g) shows that the vortex lattice is formed as an equilibrium state. 
After stopping evaporative cooling at $t=8t_0$,
the vortices decay and the condensate
size is decreased as shown in Fig. \ref{Fig:t0-5}(h).

\section{Conclusions}
We developed a simulation method for evaporative cooling of trapped BEC at finite temperatures using ZNG formalism.
In our method, the evaporative cooling process is modeled by eliminating test particles
describing the noncondensate atoms.
The simulation results successfully
reproduced
growth of BEC by evaporative cooling.
We applied our method to simulate vortex lattice formation by cooling the rotating Bose gas.
Our simulation results showed
 nucleation and formation of vortex lattice 
during the evaporative cooling 
Since we used the ZNG formalism, we could not
start with the normal gas above $T_c$ and 
go through the BEC transition. 
In order to improve the method so that it can go through the BEC transition, one will need a theory that can include the fluctuation of the condensate order parameter, such as the classical-field technique \cite{doi.org/10.1080/00018730802564254}.

\bigskip\noindent{\bf\large  Acknowledgecment  }
This work was supported by Grants-in-Aid 
(KAKENHI No. 17K14364) from the Japan Society for the Promotion of Science.
%


\begin{thebibliography}{7}
\bibitem{PhysRevLett.87.210403}
Haljan, P.C.,Coddington, I.,
Engels, P.,
Cornell, E.A.:
Phys. Rev. Lett  {\bf 87},210403
(2001).


\bibitem{C_W_Gardiner_PRL82}
Gardiner, C. W., Lee, M. D., Ballagh R. J.,  Davis M. J., and Zoller, P. :
Phys. Rev. Lett.
{\bf 82},
5266
(1998).

\bibitem{M_J_Bijlsma_PRA62}
M.~J.~Bijlsma, E.Z.,Stoof, H.T.C.:
Phys. Rev. A.
{\bf 62}, {063609}
({2000}).

\bibitem{E_arahata_JLTP2016}
Arahata, E.,
Nikuni, T.:
{J. Low Temp. Phys.}
{\bf 183},
{191}
({2016}).


\bibitem{T_ZNG_JLTP}
{Zaremba}, {E.},
{Nikuni}, {T.},
{Griffin}, {A.}:
{J. Low Temp. Phys.}
{\bf 116},
{277}
({2009}).


\bibitem{PhysRevA.66.033606}
{Jackson}, {B.},
{Zaremba}, {E.}:
{Phys. Rev. A}
{\bf 66},
{033606}
({2002}).


\bibitem{doi.org/10.1080/00018730802564254}
{Blakie}, {P.B.},
{Bradley}, {A.S.},
{Davis}, {M.J.},
{Ballagh}, {R.J.},
{Gardiner}, {C.W.}:
{Advances in Physics.}
{\bf 57},
{363}
({2008}).


\end{thebibliography}
\end{document}